\documentclass[12pt,a4paper]{article}

\usepackage{epsf}
\usepackage{latexsym,amssymb,euscript}
\usepackage[dvips]{graphicx}
\usepackage[numbers,sort&compress]{natbib}

\usepackage[italian,english]{babel}
\usepackage{indentfirst}
\usepackage[T1]{fontenc}
\usepackage{graphicx,xcolor}
\usepackage[format=hang,font=small]{caption}
\usepackage{subfig}
\usepackage{amsmath}
\usepackage{amsfonts}
\usepackage{lscape}
\usepackage{amssymb}
\usepackage{booktabs}
\usepackage{siunitx}
\usepackage{hyperref}
\usepackage{geometry}
\usepackage{slashed}

\newcommand{\beq}[1]{
\begin{equation}\label{#1}}
\newcommand{\eeq}{\end{equation}}
\newcommand{\bea}[1]{
\begin{eqnarray}\label{#1}}
\newcommand{\eea}{\end{eqnarray}}
\newcommand{\barr}{
\begin{array}}
\newcommand{\earr}{\end{array}}

\makeatletter
\def\appendix{\par\clearpage
  \setcounter{section}{0}
  \setcounter{subsection}{0}
  \@addtoreset{equation}{section}
  \def\@sectname{Appendix~}
  \def\theequation{\thesection.\arabic{equation}}
  \def\theequation{\thesection.\arabic{equation}}
  \def\thesection{\Alph{section}}}
\makeatother

\begin{document}
\begin{titlepage}

\begin{center}
{\LARGE \bf High-energy resummation in heavy-quark pair photoproduction}
\end{center}

\vskip 0.5cm

\centerline{F.G. Celiberto$^{1,2\dagger}$, D.Yu.~Ivanov$^{3,4\P}$,
B.~Murdaca$^{2\dagger}$ and A.~Papa$^{1,2\dagger}$}

\vskip .6cm

\centerline{${}^1$ {\sl Dipartimento di Fisica, Universit\`a della Calabria}}
\centerline{\sl I-87036 Arcavacata di Rende, Cosenza, Italy}
\vskip .2cm
\centerline{${}^1${\sl Istituto Nazionale di Fisica Nucleare, Gruppo collegato
      di Cosenza}}
\centerline{\sl I-87036 Arcavacata di Rende, Cosenza, Italy}
\vskip .2cm
\centerline{${}^3$ {\sl Sobolev Institute of Mathematics, 630090 Novosibirsk,
    Russia}}
\vskip .2cm
\centerline{${}^4$ {\sl Novosibirsk State University, 630090 Novosibirsk,
    Russia}}

\vskip 2cm

\begin{abstract}

We present our predictions for the inclusive production of two heavy
quark-antiqu\-ark pairs, separated by a large rapidity interval, in the
collision of (quasi-)real photons at the energies of LEP2 and of some future
electron-positron colliders.
We include in our calculation the full resummation of leading logarithms
in the center-of-mass energy and a partial resummation of the next-to-leading
logarithms, within the Balitsky-Fadin-Kuraev-Lipatov (BFKL) approach.
\end{abstract}

\vskip .5cm

$^{\dagger}${\it e-mail}:
francescogiovanni.celiberto, beatrice.murdaca, alessandro.papa
\ @fis.unical.it
  
$^{\P}${\it e-mail}: d-ivanov@math.nsc.ru

\end{titlepage}

\section{Introduction}

The high energies reached at the LHC and in possible future hadron and
electron-positron colliders represent a great chance in the search for
long-waited signals of New Physics. They offer, however, also a unique
opportunity to test the Standard Model in unprecedented kinematic ranges.
A vast class of processes can be studied at high-energy colliders,
called {\em semihard processes}, characterized by a clear hierarchy of scales,
$s \gg Q^2 \gg \Lambda_{\rm QCD}^2$, where $s$ is the squared center-of-mass
energy, $Q$ is the hard scale given by the process kinematics 
and $\Lambda_{\rm QCD}$ is the QCD mass
scale, which still represent a challenge for QCD in the high-energy limit.
Here the fixed-order perturbative description, allowed by the presence of a
hard energy scale, misses the effect of large energy logarithms, which
compensate the smallness of the coupling $\alpha_s$ and must therefore
be resummed to all orders. The theoretical framework for this resummation
is provided by the Balitsky--Fadin--Kuraev--Lipatov (BFKL) approach~\cite{BFKL},
whereby a systematic procedure has become available for resumming
all terms proportional to $(\alpha_s\ln(s))^n$, the so called
leading logarithmic approximation (LLA), and also those
proportional to $\alpha_s(\alpha_s\ln(s))^n$, the so called next-to-leading
approximation (NLA).
In both cases, within the BFKL approach, the (possibly differential) cross
section of processes falling in the domain of perturbative QCD, takes
a peculiar factorized form, whose ingredients are the impact factors
describing the transition from each colliding particle to the respective 
final state object, and a process-independent Green's function.
The BFKL Green's function obeys an integral equation, whose
kernel is known at the next-to-leading order (NLO) both for forward
scattering ({\it i.e.} for $t=0$ and color singlet in the
$t$-channel)~\cite{Fadin:1998py,Ciafaloni:1998gs} and for any fixed (not
growing with energy) momentum transfer $t$ and any possible two-gluon color
state in the $t$-channel~\cite{Fadin:1998jv,FG00,FF05}.

The phenomenological reach of the BFKL approach is limited by the number
of available impact factors. So far, only a few of them have been calculated
with next-to-leading order accuracy: i) impact factors for colliding quarks and
gluons~\cite{fading,fadinq,Cia,Ciafaloni:2000sq}, which are at the basis
of the calculation of the ii) forward jet impact factors (or jet vertices)
in exact form~\cite{bar1,bar2,Caporale:2011cc} or in the small-cone
approximations~\cite{Ivanov:2012ms,Colferai:2015zfa} and of the iii) forward
hadron impact factors~\cite{Ivanov:2012iv}, iv) impact factor for the
$\gamma^*$ to light vector meson transition at leading twist~\cite{IKP04},
v) impact factor for the $\gamma^*$ to $\gamma^*$
transition~\cite{gammaIF,Balitsky2012}.

Jet vertices have extensively been used to produce with NLA accuracy
a number of predictions~\cite{Colferai:2010wu,Angioni:2011wj,Caporale:2012ih,Ducloue:2013wmi,Ducloue:2013bva,Caporale:2013uva,Ducloue:2014koa,Caporale:2014gpa,Ducloue:2015jba,Caporale:2015uva,Celiberto:2015yba,Celiberto:2016ygs,Chachamis:2015crx} for the Mueller-Navelet jet production process at the LHC, resulting
in nice agreement with experimental determinations~\cite{Khachatryan:2016udy}.
The same vertices enter the calculation of several observables in the inclusive
production of three and four jets, separated in rapidity, at the LHC~\cite{Caporale:2015vya,Caporale:2015int,Caporale:2016soq,Caporale:2016xku,Caporale:2016zkc}.

The forward hadron impact factors were recently used to
calculate the cross section and some azimuthal correlations in the inclusive
production of two identified hadrons composed of light quarks and separated in
rapidity~\cite{Celiberto:2016hae,Celiberto:2017ptm} which could also be
studied at the LHC.

The impact factor for the $\gamma^*$ to light vector meson transition enters
the imaginary part of the cross section for the exclusive production of two
light vector mesons in the collision of two highly virtual
photons~\cite{IP06,IP07,Segond:2007fj,Enberg:2005eq,Pire:2005ic}, which
could be considered in future linear colliders.

The $\gamma^*$ to $\gamma^*$ impact factor is the ingredient for the
$\gamma^* \gamma^*$ total cross section, which is considered to be the
gold-plated channel for the manifestation of the BFKL dynamics. A number
of predictions for this cross section were built, with partial inclusion
of NLA BFKL effects~\cite{Brodsky:2002ka,Brodsky:1998kn,Caporale2008,Zheng}
and with full NLA accuracy~\cite{Chirilli2014,Ivanov:2014hpa}, whose
comparison with the only available data from LEP2 cannot be conclusive due to
the relatively small center-of-mass energy and the limiting accuracy of LEP2
experiment.

\begin{figure}[h]
\centering
\includegraphics[scale=0.75]{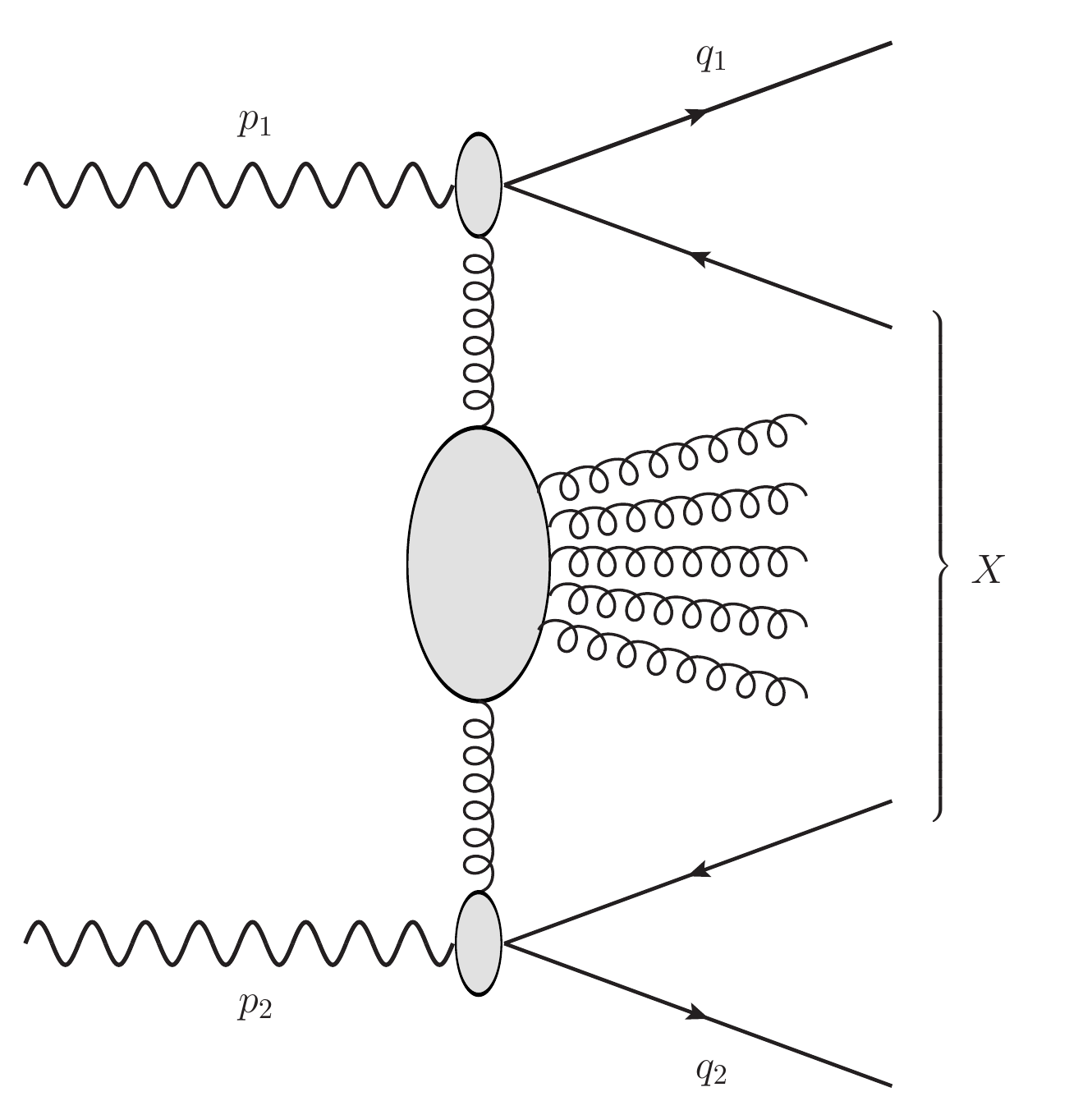}
\caption[]{Diagrammatic representation of the heavy-quark pair photoproduction
  in the case when a heavy quark with transverse momentum $q_1$ ($q_2$) from
  the upper (lower) vertex is tagged.}
\label{fig:process}
\end{figure}

In this paper we introduce another process which could serve as a probe of
BFKL dynamics: the inclusive production of two heavy quark-antiquark pairs,
separated in rapidity, in the collision of two real (or quasi-real) photons,
\beq{process}
\gamma(p_1) + \gamma(p_2)  \longrightarrow Q(q_1) 
\ + X \ + Q(q_2) \;,
\eeq
where $Q$ here stands for a charm/bottom quark {\em or} antiquark. In
Fig.~\ref{fig:process} we present a schematic representation of this process,
in the case when the tagged objects are a heavy quark with momentum $q_1$,
detected in the fragmentation region of the photon with momentum $p_1$,
and a heavy quark with momentum $q_2$, detected in the fragmentation
region of the photon with momentum $p_2$. This process can be studied either
at electron-positron or in nucleus-nucleus colliders via collisions of two
quasi-real photons. In this first exploratory study, we will focus on the
case of electron-positron colliders and will adopt the equivalent photon
approximation (EPA) to parametrize the photon flux emitted by the colliding
electrons and positrons. The main aim is to show that sensible predictions
can be built, within the BFKL approach with NLA, which can be compared
with experimental results. For the sake of definiteness, we will consider
the center-of-mass energies of LEP2 and of the CLIC future collider.

The totally inclusive two heavy-quark pair production process has much in
common with the above discussed inclusive interaction of two virtual photons
(the $\gamma^*\gamma^*$ total cross section). Here the large values of masses
of the produced heavy quarks play the role of hard scale, similar to the role
that large photon virtualities play in $\gamma^*\gamma^*$ interactions.
It is interesting to note that just this observable, the total inclusive cross
section for two heavy-quark pairs photoproduction, was calculated first in QCD
within BFKL resummation method, see the paper by I.~Balitsky and L.~Lipatov
in~\cite{BFKL}. Despite the fact that the BFKL resummation gives formally
a finite result for this total cross section, it does not represent an
observable that can be directly confronted with the experiment. Indeed, in
order to be sure that two heavy-quark pairs are produced in the event, one
needs to detect at least one of the heavy quarks in each quark pair. The other
reason for the tagging of two heavy quarks is that the knowledge of their
momenta (their rapidities) allows one to keep control on the energy of the
collision of two quasi-real photons in $e^+e^-$ experiments. 
In our present study we restrict/fix the momenta of these two tagged quarks as
if they were true final states.  As a further step, one needs to include into
the theoretical analysis the heavy-quark fragmentation describing the tagging
procedure of heavy quarks in the particular experiment. 

An attractive idea is to consider also similar experiments which assume the
detection of the pair of heavy quarks, separated by a large rapidity interval,
in photon-photon interactions via ultra-peripheral (UPC) nucleus-nucleus
collisions at the LHC. 
In~\cite{Goncalves:2003jv} the total cross section for the production of two
heavy-quark pairs in such collisions was estimated in the LO BFKL approach at a
sizeable value. However the kinematics of experiments with a tagged pair of
heavy quarks separated by the rapidity interval of a few units requires rather
large energies of the colliding quasi-real photons. Unfortunately at the LHC
the energies of such photon-photon interactions in the UPC heavy nucleus-nucleus
collisions fall in the kinematic range where the quasi-real photon fluxes from
the colliding heavy nuclei are greatly suppressed due to electromagnetic
nuclear form factors, and therefore such experiments look not feasible.

The paper is organized as follows: in Section~2 we explain the theoretical
setup of our calculation; in Section~3 we present our results for the cross
sections azimuthal angle correlations in dependence on the rapidity interval
between the tagged heavy quarks; in Section~4 we discuss our results and
draw conclusions.

\section{Theoretical setup}

\begin{figure}[h]
\centering
\includegraphics{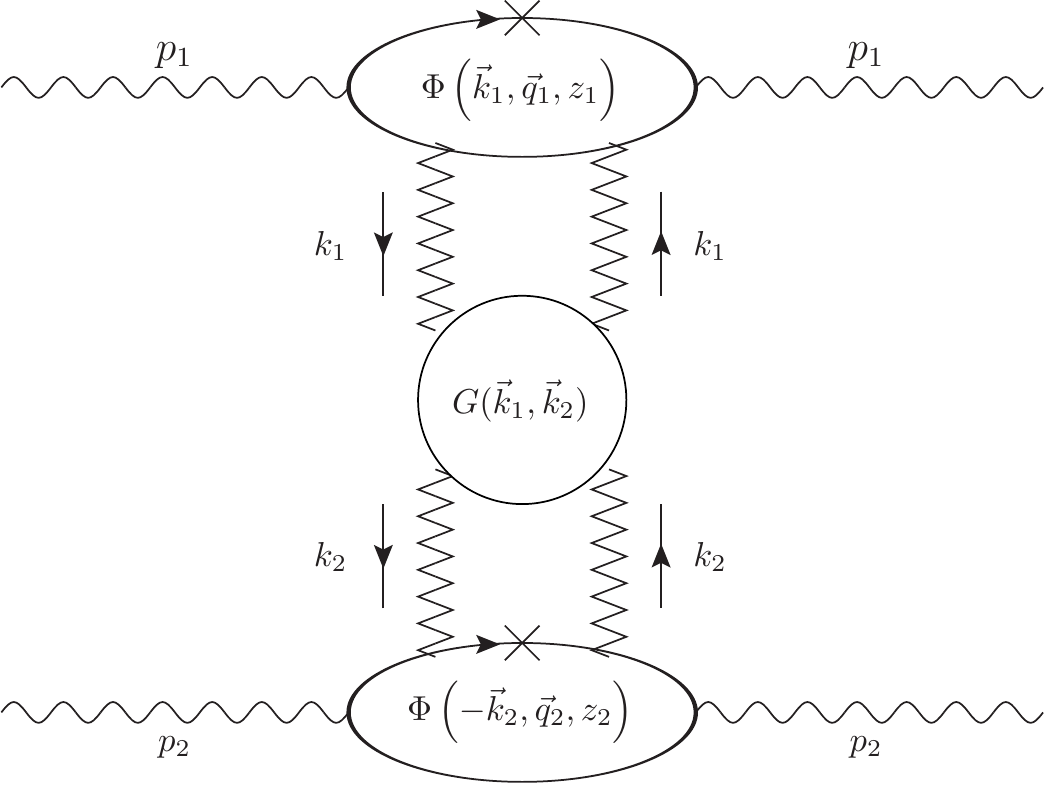}
\caption[]{Schematic representation of the BFKL factorization for the
  process under consideration.}
\label{bfkl}
\end{figure}

For the process under consideration, given in Eq.~(\ref{process}), we plan to
construct the cross section, differential in some of the kinematic variables
of the tagged heavy quark or antiquark, and some azimuthal correlations
between the tagged fermions. In the BFKL approach the cross section takes
a factorized form, schematically represented in Fig.~\ref{bfkl}, given by
the convolution of the impact factors for the transition from a (quasi-)real
photon to a heavy quark-antiquark pair (the upper and lower ovals in
Fig.~\ref{bfkl}, labeled by $\Phi$) with the BFKL Green's function $G$.
The crosses in Fig.~\ref{bfkl} denote the tagged quarks, whose momenta are
not integrated over in getting the expression for the cross section.

In our calculation we will partially include NLA resummation effects, by
taking the BFKL Green's function in the NLA, while keeping the impact factors
at the leading order, since their next-to-leading order corrections are not
yet known.

\subsection{The impact factor}

The (differential) impact factor for the photoproduction of a heavy
quark pair reads 
\begin{equation}
  d{\Phi}=\frac{\alpha\alpha_s e_Q^2}{\pi}\left[ m^2R^2+\vec P^2\left(z^2
    +\overline z^2\right)\right]d^2q\ dz\, ,
\end{equation}
where $R$ and $\vec P$ read
\begin{equation}
  R=\frac{1}{m^2+\vec q^{\, \,2}}-\frac{1}{m^2+(\vec q-\vec k)^2}\;,
  \;\;\;\;\;
  \vec P=\frac{\vec q}{m^2+\vec q^{\, \, 2}}+\frac{\vec k-\vec q}
       {m^2+(\vec q-\vec k)^2}\, .
\end{equation}
Here $\alpha$ and $\alpha_s$ denote the QED  and  QCD couplings, $e_Q$ denotes
the electric charge of the heavy quark, $m$ stands for the heavy-quark mass, $z$
and ${\overline z}\equiv 1-z$ are the longitudinal fractions
of the quark and antiquark produced in the same vertex and $\vec k$, $\vec q$, $\vec k-\vec q$ represent the
transverse momenta with respect to the photons collision axis of the Reggeized
gluon, the produced quark and antiquark, respectively.
The details of the derivation of this result may be found, for instance,
in~\cite{Ginzburg:1996vq}. 
Such impact factor differs only by the coupling and overall normalization from
the similar QED quantity known since long and used in the calculations of the
lepton-pair production. 

In the following we will need the projection of the impact factors onto the
eigenfunctions of the leading-order BFKL kernel, to get their so called
$\left(n,\nu\right)$-representation. We get
\[
v_{R^2}\equiv\int\frac{d^2k}{\pi\sqrt 2}\left(k^2\right)^{i\nu-3/2}e^{in\vartheta}R^2
\]
\[
=\frac{1}{\sqrt2}\frac{\Gamma\left(\frac{1}{2}+\frac{n}{2}-i\nu \right)
  \Gamma\left(\frac{1}{2}+\frac{n}{2}+i\nu \right)
  \left( \vec q^{\, \, 2}\right)^{\frac{n}{2}}e^{in\varphi}\left(\frac{1}{2}
  +\frac{n}{2}-i\nu \right)}{\Gamma\left(n+1\right)
  \left(m^2+\vec q^{\, \, 2} \right)^{\frac{5}{2}+\frac{n}{2}-i\nu}
  \left(\frac{n}{2}+i\nu-\frac{1}{2} \right)}
\]
\[
\times\left[\left(\frac{3}{2}+\frac{n}{2}-i\nu\right)\,
  _2F_1\left(\frac{n}{2}-\frac{1}{2}+i\nu,\, \frac{5}{2}+\frac{n}{2}-i\nu,
  \, 1+n,\, \zeta \right) \right.
  \]
\beq{proj_R}
-\left.2\, _2F_1\left(\frac{n}{2}-\frac{1}{2}+i\nu,\, \frac{3}{2}+\frac{n}{2}
-i\nu,\, 1+n,\, \zeta \right) \right]
\eeq
\[
\equiv e^{in\varphi} \, c_{R^2}(n,\nu,\vec q^{\:2})
\]
and
\[
v_{\vec P^{\, 2}} \equiv
\int\frac{d^2k}{\pi\sqrt2}\left(k^2\right)^{i\nu-3/2}e^{in\vartheta}\vec P^{\, 2}
\]
\[
=\frac{1}{\sqrt2}\frac{\Gamma\left(\frac{1}{2}+\frac{n}{2}-i\nu \right)
  \Gamma\left(\frac{1}{2}+\frac{n}{2}+i\nu \right)\left( \vec q^{\, \, 2}
  \right)^{\frac{n}{2}}e^{in\varphi}\left(\frac{1}{2}+\frac{n}{2}-i\nu \right)}
     {\Gamma\left(1+n\right)\left(m^2+\vec q^{\, \, 2} \right)^{\frac{3}{2}
         +\frac{n}{2}-i\nu}\left(-\frac{1}{2}+\frac{n}{2}+i\nu \right)}
\]
\[
\times \left\lbrace\, \left[1 + 2i\nu\,(1-\zeta) \right]
{_2}F_1\left(\frac{n}{2}-\frac{1}{2}+i\nu,\, \frac{3}{2}+\frac{n}{2}-i\nu,
\, 1+n,\, \zeta \right) \right. 
\]
\beq{proj_P}
+ \left.
 \frac{\nu^2-i\nu+\frac{3}{4}+\frac{n^2}{4}}{i\nu-\frac{1}{2}-\frac{n}{2}}\, 
{_2}F_1\left(\frac{n}{2}-\frac{1}{2}+i\nu,\, \frac{1}{2}+\frac{n}{2}-i\nu,
\, 1+n,\, \zeta \right)  \right\rbrace
\eeq
\[
\equiv e^{in\varphi} \, c_{\vec P^{\,2}}(n,\nu,\vec q^{\:2})\;,
\]
where $\zeta \equiv \frac{\vec q^{\, \, 2}}{m^2+\vec q^{\, \, 2}}$;
the azimuthal angles $\vartheta$ and $\varphi$ are defined as
$\cos\vartheta\equiv k_x/|\vec k|$ and $\cos\varphi\equiv q_x/|\vec q|$.

\subsection{Kinematics of the process}

For the tagged quark momenta we introduce the standard Sudakov decomposition,
using as light-cone basis the momenta $p_1$ and $p_2$ of the colliding photons,
\begin{equation}
q=z p_1+\frac{m^2+\vec q^{\, \,2}}{z W^2}p_2+q_{\perp}\, ,
\end{equation}
with $W^2=\left(p_1+p_2\right)^2=2p_1p_2=4E_{\gamma_1}E_{\gamma_2}$;
$p_1=\frac{E_{\gamma_1}}{2}\left(1,\vec 0,1\right)$ and
$p_2=\frac{E_{\gamma_2}}{2}\left(1,\vec 0,-1\right)$, so that 
\[
2qp_2=W^2z=2E_{\gamma_2} \left(E+q_{\parallel}\right)\ ,
\]
\[
2qp_1=\frac{m^2+\vec q^{\, \, 2}}{z}=2E_{\gamma_1}\left(E-q_{\parallel}\right)
\; ;
\]
here $q=\left(E,\vec q,q_\parallel\right)$ and  the rapidity can be expressed as
\[
y=\frac{1}{2}\ln\frac{E+q_{\parallel}}{E-q_{\parallel}}
=\ln\frac{2E_{\gamma_1}z}{\sqrt{m^2+\vec q^{\, \, 2}}} \;.
\]
Therefore for the rapidities of the two tagged quarks in our process we have 
\[
y_1=\ln\frac{2E_{\gamma_1}z_1}{\sqrt{m^2+\vec q_1^{\, \, 2}}}\;\;\;
\text{and} \;\;\; y_2=-\ln\frac{2E_{\gamma_2}z_2}{\sqrt{m^2+\vec q_2^{\, \, 2}}}\;,
\]
whence their rapidity difference is
\[
\Delta Y \equiv y_1-y_2=\ln\frac{W^2z_1z_2}
{\sqrt{\left(m^2+\vec q_1^{\, \, 2}\right)\left( m^2+\vec q_2^{\, \, 2}\right)}}\;.
\]
For the semihard kinematic we have the requirement   
\[
\frac{W^2}{\sqrt{\left(m^2+\vec q_1^{\,\,2}\right)
    \left(m^2+\vec q_2^{\,\,2}\right)}}=\frac{e^{\Delta Y}}{z_1z_2}\gg1\;,
\]
therefore we will consider the kinematic when $\Delta Y\geq\Delta_0\sim 1\div
2$.

In what follows we will need a cross section differential in the rapidities of
the tagged quarks, therefore we have to make the following change of variables:
\[
z_1\to y_1=\ln\frac{2E_{\gamma_1}z_1}{\sqrt{m^2+\vec q_1^{\, \, 2}}}\;,
\;\;\; dy_1=\frac{dz_1}{z_1}\;,
\]
\[
z_2\to y_2=-\ln\frac{2E_{\gamma_2}z_2}{\sqrt{m^2+\vec q_2^{\, \, 2}}}\;,
\;\;\; dy_2=-\frac{dz_2}{z_2}\;,
\]
which implies
\[
dz_1dz_2=\frac{e^{y_1-y_2}\sqrt{m^2+\vec q^{\, \, 2}_1}
  \sqrt{m^2+\vec q^{\, \, 2}_2}}{W^2}dy_1dy_2 \;.
\] 

\subsubsection{The BFKL cross section and azimuthal coefficients}

Similarly to the Mueller-Navelet jet and the dihadron production processes
(see Refs.~\cite{Caporale:2012ih,Celiberto:2017ptm}), the 
differential cross section for the inclusive production of a pair of heavy
quarks separated in rapidity (a ``diquark'' system in what follows) can
be cast in the following form:
\[
\frac{d\sigma}{dy_1 dy_2 d|\vec q_1| d|\vec q_2| d\phi_1 d\phi_2}
= \frac{1}{(2\pi)^2} \left[{\cal C}_0 +2\sum_{n=1}^\infty \cos(n\varphi) {\cal C}_n
\right]\;,
\]
where $\varphi = \varphi_1-\varphi_2-\pi$, while ${\cal C}_0$ gives the
cross section averaged over the azimuthal angles $\varphi_{1,2}$ of the 
produced quarks and the other coefficients ${\cal C}_n$ determine the
distribution of the relative azimuthal angle between the two quarks.

The expression for the ${\cal C}_n$ coefficient is the following
($q_{1,2}\equiv |\vec q_{1,2}|$):
\[
{\cal C}_n = \frac{q_1 q_2 \, \sqrt{m_1^2+q^{ 2}_1} \sqrt{m_2^2+q^{ 2}_2}}
{W^2} e^{\Delta Y}
\]
\[
\times  \int d\nu\left(\frac{W^2}{s_0} \right)^{{\overline\alpha_s}
\left(\mu_R\right)\chi\left(n,\nu\right)+\overline\alpha_s^2\left(\mu_R\right)
\left(\bar\chi\left(n,\nu\right)+\frac{\beta_0}{8N_c}\chi\left(n,\nu\right)
\left(-\chi\left(n,\nu\right)+\frac{10}{3}+2\ln\frac{\mu_R^2}{\sqrt{s_1s_2}}\right)
\right)}
\]
\[
\times \alpha_s^2\left(\mu_R\right) c_1(n,\nu,\vec q_1^{\:2},z_1)
c_2(n,\nu,\vec q_2^{\:2},z_2)
\]
\[
\times \left\lbrace1+{\overline\alpha_s}\left(\mu_R\right)
\left(\frac{\bar c_1^{(1)}}{c_1}+\frac{\bar c_2^{(1)}}{c_2}\right)
+{\overline\alpha_s}\left(\mu_R\right)\frac{\beta_0}{2N_c}
\left(\frac{5}{3}+\ln\frac{\mu_R^2}{s_1s_2} +f\left(\nu\right)\right) \right.
\]
\beq{Cn}
\left. + \overline\alpha_s^2 (\mu_R)\, \ln\left({\frac{W^2}{s_0}}\right)
\,\frac{\beta_0}{4N_c}\, \chi\left(n,\nu\right) f(\nu) \right\rbrace\;,
\eeq
where 
\[
\chi\left(n,\nu\right)=2\psi\left( 1\right)-\psi\left(\frac{n}{2}
+\frac{1}{2}+i\nu \right)-\psi\left(\frac{n}{2}+\frac{1}{2}-i\nu \right)
\]
are the eigenvalues of the leading-order BFKL kernel, with
$\psi(x)=\Gamma'(x)/\Gamma(x)$, and
\[
\beta_0=\frac{11}{3} N_c - \frac{2}{3}n_f
\]
is the first coefficient of the QCD $\beta$-function, responsible for
running-coupling effects. The function
$f\left(\nu\right)$ is defined by 
\[
i\frac{d}{d\nu}\ln\frac{c_1}{c_2}=2\left[f\left(\nu\right)
  -\ln\left(\sqrt{s_1s_2}\right)\right]\;,
\]
with $s_i$, $i=1,2$ being the hard scales in our two-tagged-quark process,
which we chose equal to $m_i^2 + \vec q_i^{\:2}$\;, and
\[
c_1(n,\nu,\vec q_1^{\:2},z_1) = \frac{\alpha e_{Q_1}^2}{\pi}
\left[m_1^2c_{R^2}(n,\nu,\vec q_1^{\:2})+\left(z_1^2+\overline z_1^2\right)
  c_{\vec P^2}(n,\nu,\vec q_1^{\:2})\right]\;,
\]
\[
c_2(n,\nu,\vec q_2^{\:2},z_2) = \frac{\alpha e_{Q_2}^2}{\pi}
\left[m_2^2c^*_{R^2}(n,\nu,\vec q_2^{\:2})+\left(z_2^2+\overline z_2^2\right)
  c^*_{\vec P^2}(n,\nu,\vec q_2^{\:2})\right]\;,
\]
\[
\frac{\bar c_1^{(1)}}{c_1}+\frac{\bar c_2^{(1)}}{c_2}=\chi\left(n,\nu\right)
\ln\frac{s_0}{\sqrt{\left(m_1^2+\vec q^{\, \, 2}_1\right)
\left(m_2^2+\vec q^{\, \, 2}_2\right)}}\;.
\]
The scale $s_0$ can be arbitrarily chosen, within NLA accuracy; in our
calculation we made the choice $s_0=\sqrt{s_1 s_2}$. Equation~(\ref{Cn}) is
written for the general case when two heavy quarks of different flavors with
masses $m_1$ and $m_2$ are detected.

\subsection{The $e^+e^-$ cross section}

To pass from the photon-initiated process to the one initiated by
$e^+e^-$ collisions, we must take into account the flux of quasi-real photons
$dn$ emitted by each of the two colliding particles,
\[
d\sigma_{e^+e^-}=dn_1dn_2d\sigma_{\gamma\gamma}\;,
\]
with
\beq{dn}
dn=\frac{\alpha}{\pi}\frac{dx}{x}\left[1-x+\frac{x^2}{2}
  -\frac{m_e^2\left(1-x\right)x^2}{\vec q^{\, \, 2}+m_e^2x^2}\right]
\frac{d\vec q^{\, \, 2}}{\vec q^{\, \, 2}+m_e^2x^2}\;,
\eeq
where $x=\frac{\omega}{E_e}$ is the fraction of the electron (positron) energy
carried by the photon and $\vec q$ denotes now the transverse component
of the photon momentum. The emission angle is $\theta\approx\frac{q_{\perp}}
{E_e\left(1-x\right)}$ and we will consider the antitag experiment, so that
$\theta\leq\theta_0$ whence $(\vec q)_{\rm max}=E_e\left(1-x\right)\theta_0$.

Integrating in Eq.~(\ref{dn}) over $\vec q^{\, \, 2}$, we get
\[
dn=\frac{\alpha}{\pi}\frac{dx}{x}\left[\left(1-x+\frac{x^2}{2}\right)
  \ln\left(\frac{E_e^2\theta_0^2\left(1-x\right)^2+m_e^2x^2}{m_e^2x^2}\right)
  -(1-x)\right]\;,
\]
where some terms ${\cal O}(m_e^2/E_e^2)$ were neglected.

Therefore the general expression for our observables is 
\[
\frac{d\sigma_{e^+e^-}}{d\left(\Delta Y\right)}=
\int d q_1\int d q_2\int_{-y_{\rm max}^{(1)}}^{y_{\rm max}^{(1)}} dy_1
\int_{-y_{\rm max}^{(2)}}^{y_{\rm max}^{(2)}} dy_2\, \delta\left(y_1-y_2-\Delta Y\right)
\]
\beq{final}
\times \int_{e^{-\left(y_{\rm max}^{(1)}-y_1\right)}}^{1}\frac{dn_1}{dx_1}dx_1
\int_{e^{-\left(y_{\rm max}^{(2)}+y_2\right)}}^{1}\frac{dn_2}{dx_2}dx_2\,
d\sigma_{\gamma\gamma}\;,
\eeq
with $y_{\rm max}^{(1)}=\ln\sqrt{\frac{s}{m_1^2+\vec q^{\, \, 2}_1}}$ and
$y_{\rm max}^{(2)}=\ln\sqrt{\frac{s}{m_2^2+\vec q^{\, \, 2}_2}}$.

In order to give predictions to be confronted with experiment, we have to
integrate our fully differential cross section over some range of the tagged
quarks transverse momenta. In what follows we label such \emph{integrated}
coefficients with $C_n$.
 
\subsection{The ``box'' $q\bar{q}$ cross section}

\begin{figure}[h]
\centering
\includegraphics[scale=0.75]{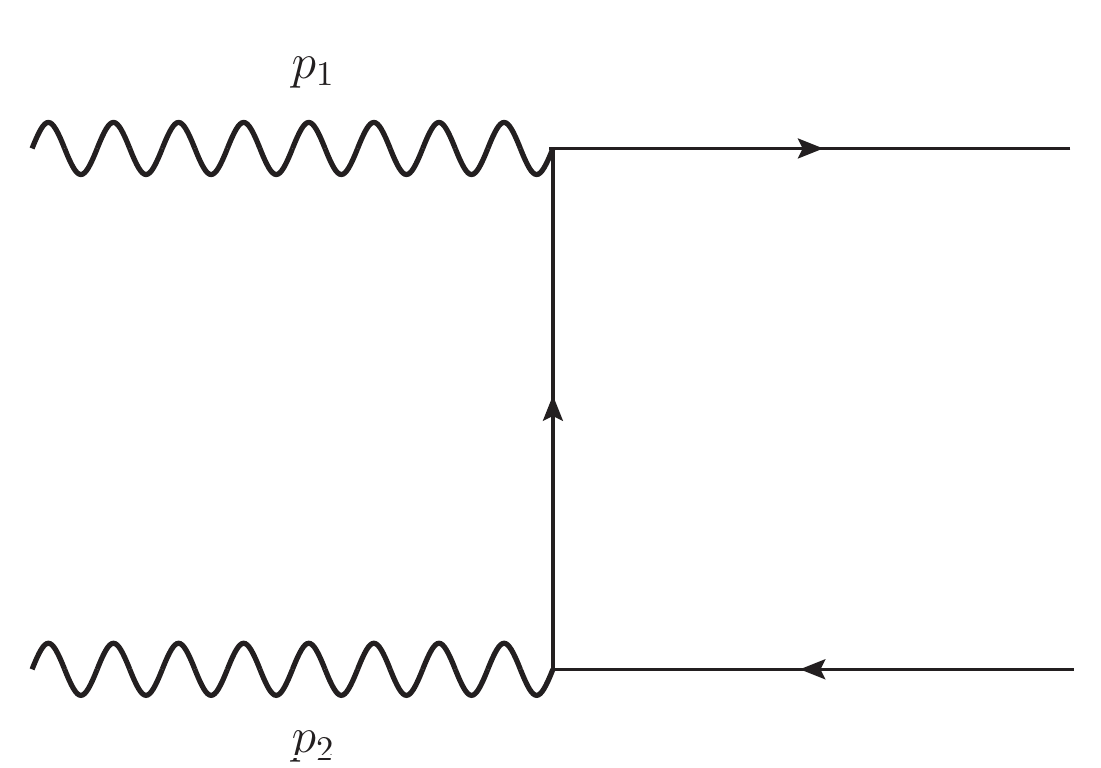}
\caption[]{One of the two Feynman diagrams contributing at the lowest
  order to the $q\bar{q}$ photoproduction. The other one is the diagram
  with crossed fermionic lines.}
\end{figure}

As a background contribution for the case when a quark and an antiquark of the
same flavor are detected, we have to consider the lowest-order QED cross
section for the production of a heavy quark and antiquark in photon-photon
collisions. In our notations the corresponding $e^+ e^-$ cross section reads 
\[
\frac{d\sigma_{ee}}{d(\Delta Y)}=\int_0^{\frac{s_{ee}}{2(1+\cosh(\Delta Y))}-m^2}
\frac{d q^2}{(m^2+q^2)^2}\frac{2\pi\alpha^2e_q^4 N_c}{\left(1+\cosh(\Delta Y)
  \right)^2}
\]
\[
\times \left[\frac{\cosh(\Delta Y)}{2}+\frac{m^2}{m^2+q^2}
  -\left( \frac{m^2}{m^2+q^2}\right)^2\right]
\]
\beq{box}
\times\left(\frac{\alpha}{\pi}\right)^2\left[f(y)\left(\ln\left(\frac{\Lambda^2}
  {m_e^2y}\right)-1 \right)^2-\frac{1}{3}\left(\ln\frac{1}{y}\right)^3\right]\;,
\eeq
where
\[
y=\frac{w^2}{s_{ee}}=\frac{2\left(1+\cosh(\Delta Y)\right)
  \left( m^2+q^2\right)}{s_{ee}} \;,
\]
with
\[
f(y)=\left(1+\frac{y}{2}\right)^2\ln\frac{1}{y}-\frac{1}{2}
\left(1-y\right)\left(3+y\right)
\]
and $\Lambda\simeq m^2$.

In the case when the heavy quarks of different flavor, two quarks or
two antiquark of the same flavor, are detected, the ``box'' mechanism does
not represent, of course, a background channel.

\section{Numerical analysis}

\subsection{Results}

In this Section we present our results for the dependence on the rapidity
separation between the two tagged quarks, $\Delta Y=y_1-y_2$, of the
$\varphi$-averaged cross section $C_0$ and of the ratios $R_{10}\equiv
C_1/{C}_0$ and $R_{20}\equiv C_2/{C}_0$ ratios. We consider here only the
case of charm quark and fix the mass $m$ at the value 1.2 GeV/$c^2$.

Introducing some reasonable kinematic cuts, we integrate the quark
transverse momenta in the symmetric range $q_{\rm min} < q_{1,2} < q_{\rm max}$.
We fix $q_{\rm max}$ to 10 GeV and consider below the three cases
$q_{\rm min} = 0$, 1, 3 GeV. We fix the center-of-mass energy
to $\sqrt{s} = 200$ GeV, typical of LEP2 analyses, and study the behavior of
our observables in the rapidity range $1 < \Delta Y < 6$. For comparison, we
give predictions of $C_0$ and $R_{10}$ also at $\sqrt{s} = 3$ TeV,
characteristic of the future $e^+e^-$ CLIC linear accelerator. In the last
case, we allow for a larger rapidity interval between the two quarks,
\emph{i.e.} $1 < \Delta Y < 11$. 
We fix the maximum for the lepton emission angle $\theta_0 = 0.0835$, which
is inside of the acceptance range of the OPAL forward
detectors~\cite{Ahmet:1990eg,Abbiendi:2002tp}. All calculations are done in the
$\overline{\rm MS}$ scheme. 

Pure LLA and NLA BFKL predictions, together with the ``box'' $q\bar{q}$
calculation of $C_0$ for $q_{\rm min} = 0$ GeV and $\sqrt{s} = 200$ GeV, are shown
respectively in Table~\ref{tab:C0-200GeV} and~\ref{tab:C0-3TeV}.

\begin{figure}[tb]
\centering

   \includegraphics[scale=0.40,clip]{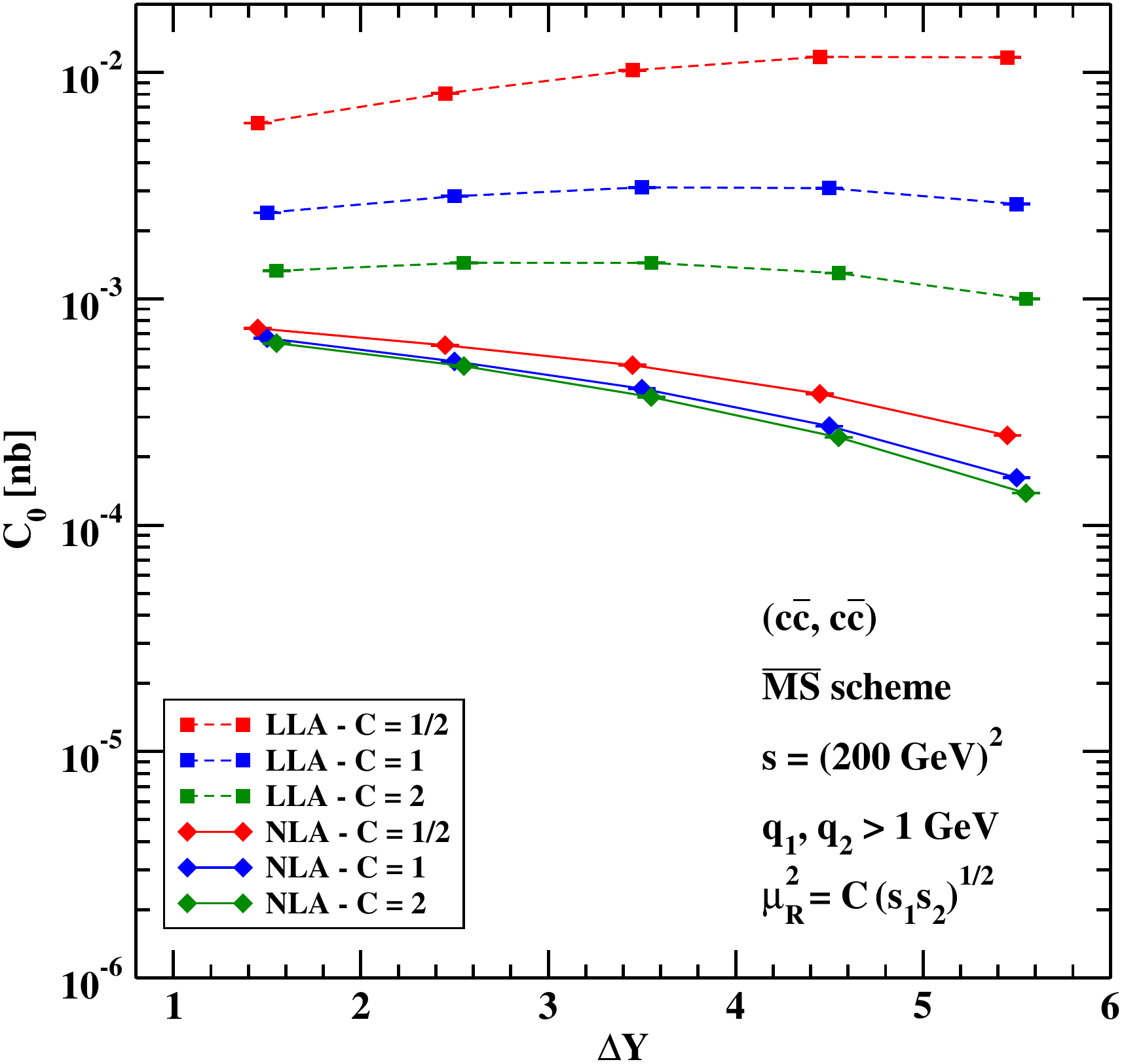}
   \includegraphics[scale=0.40,clip]{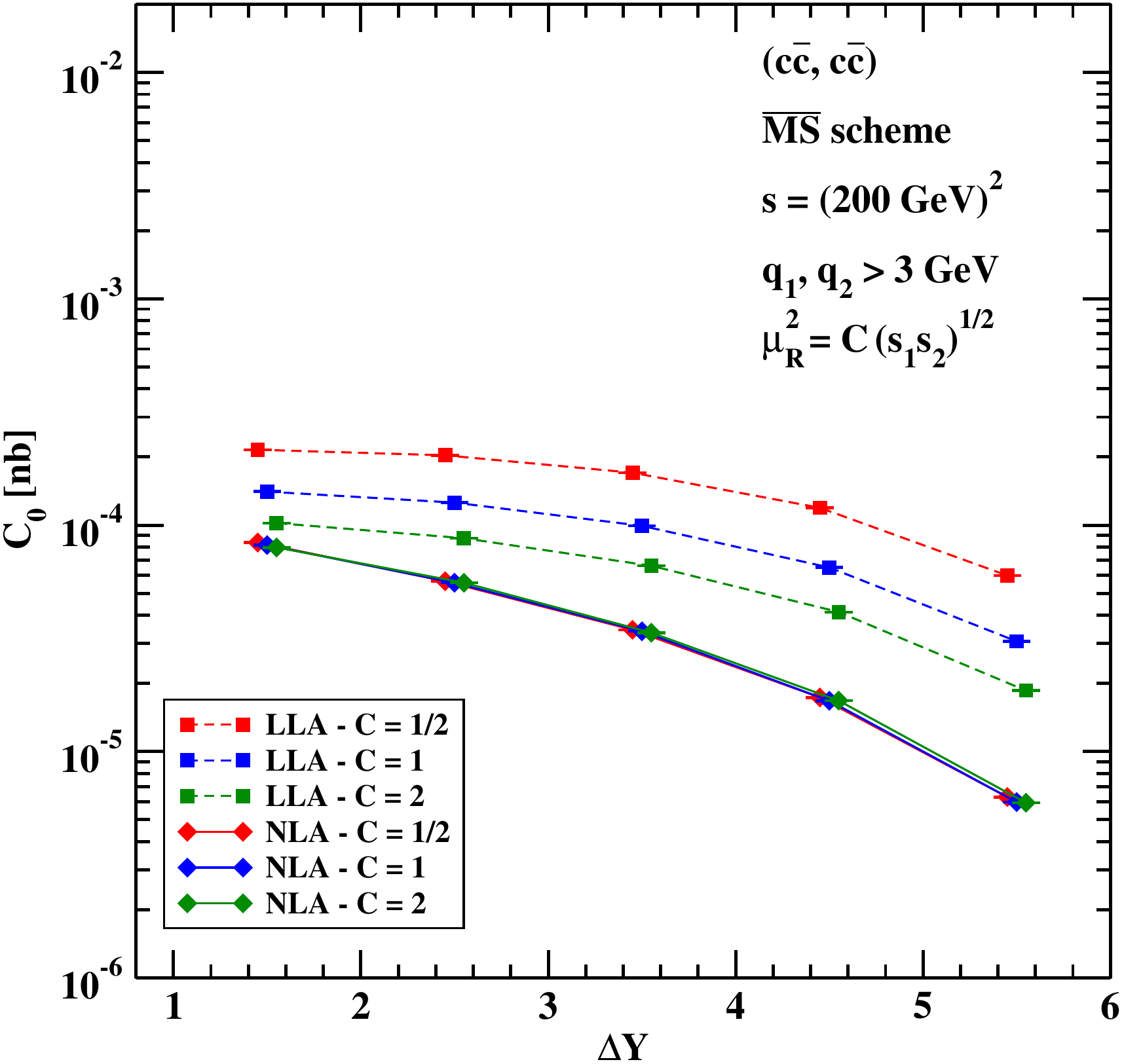}

   \includegraphics[scale=0.40,clip]{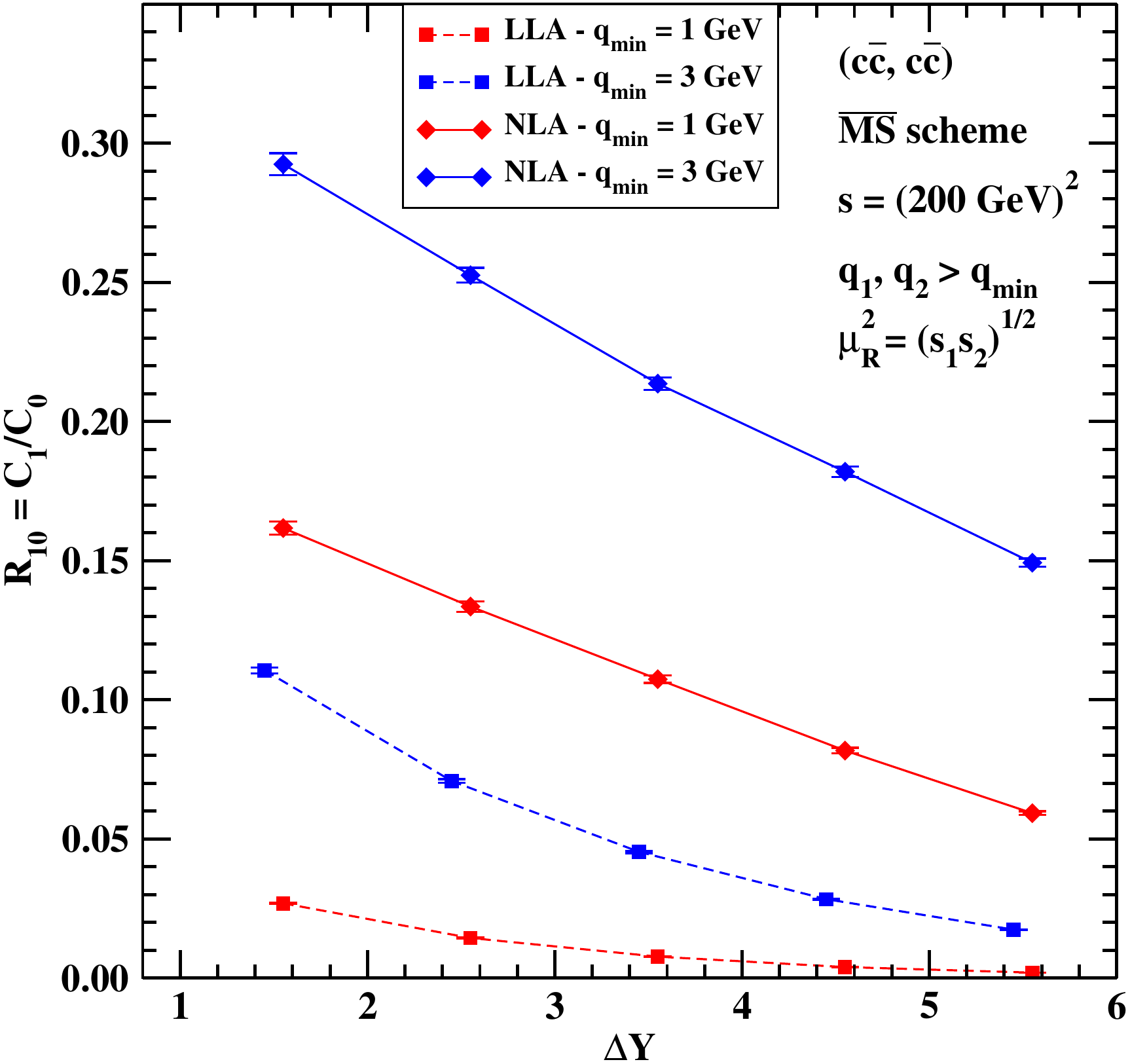}
   \includegraphics[scale=0.40,clip]{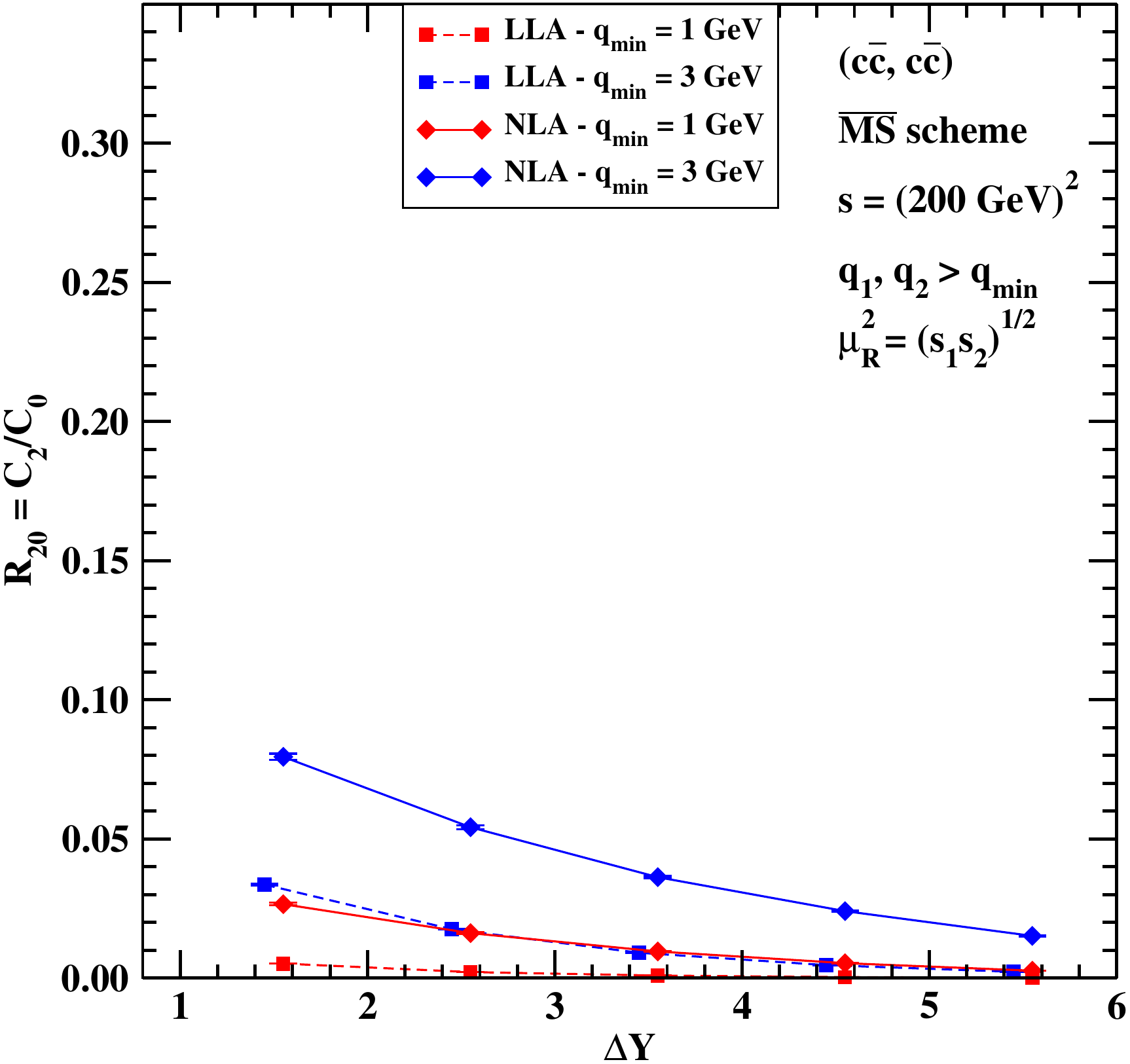}

   \caption{$\Delta Y$-dependence of $C_0$, $R_{10}$, and $R_{20}$ for
     $q_{\rm min} = 1$, 3 GeV, $\sqrt{s} = 200$ GeV, and for different values of
     $C = \mu_R^2/\sqrt{s_1 s_2}$, with $s_{1,2} = m_{1,2}^2 + q_{1,2}^2$. Data
     points have been slightly shifted along the horizontal axis for the sake
     of readability.}
\label{fig:200GeV}
\end{figure}

\begin{figure}[tb]
\centering

   \includegraphics[scale=0.40,clip]{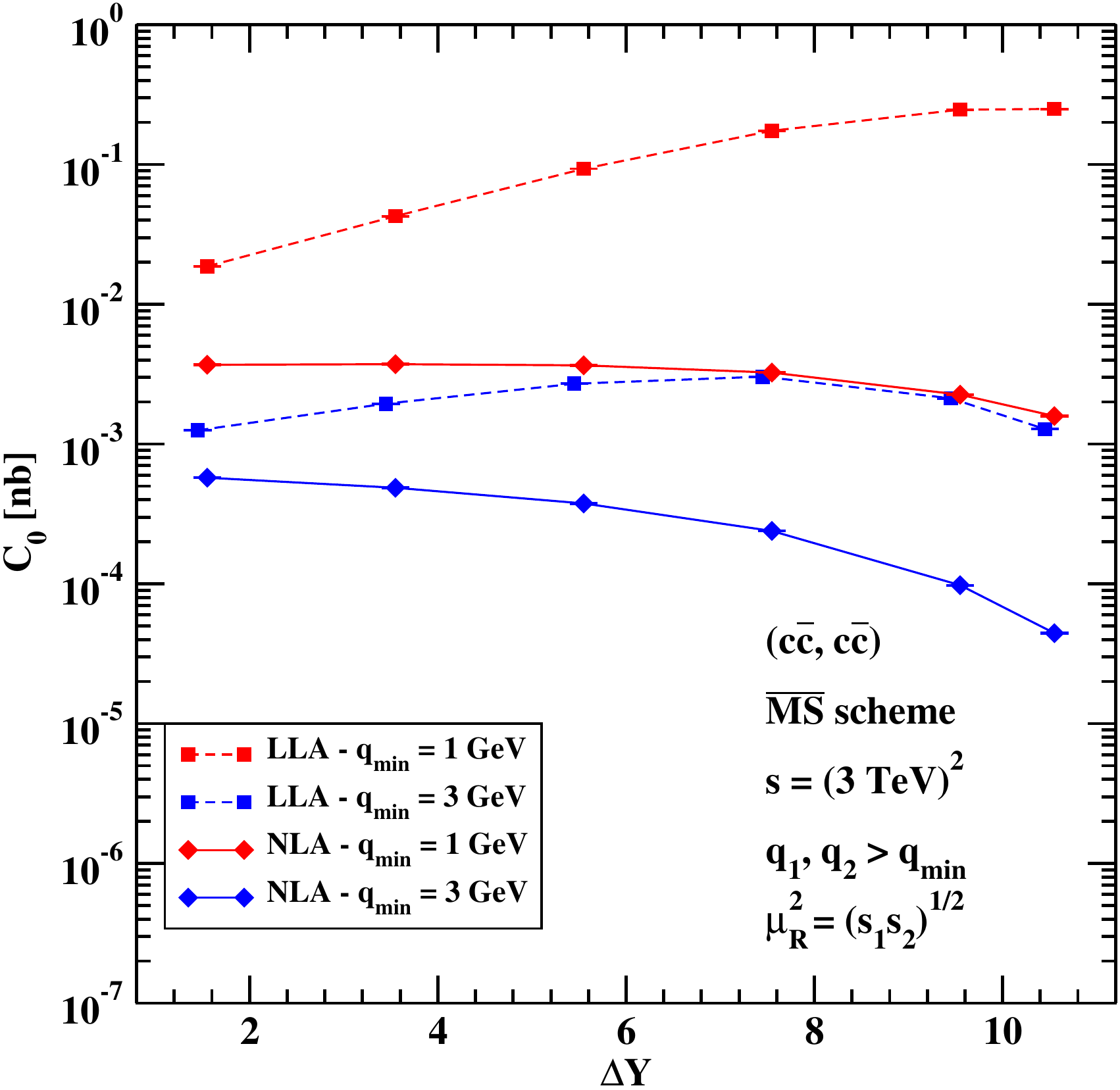}
   \includegraphics[scale=0.40,clip]{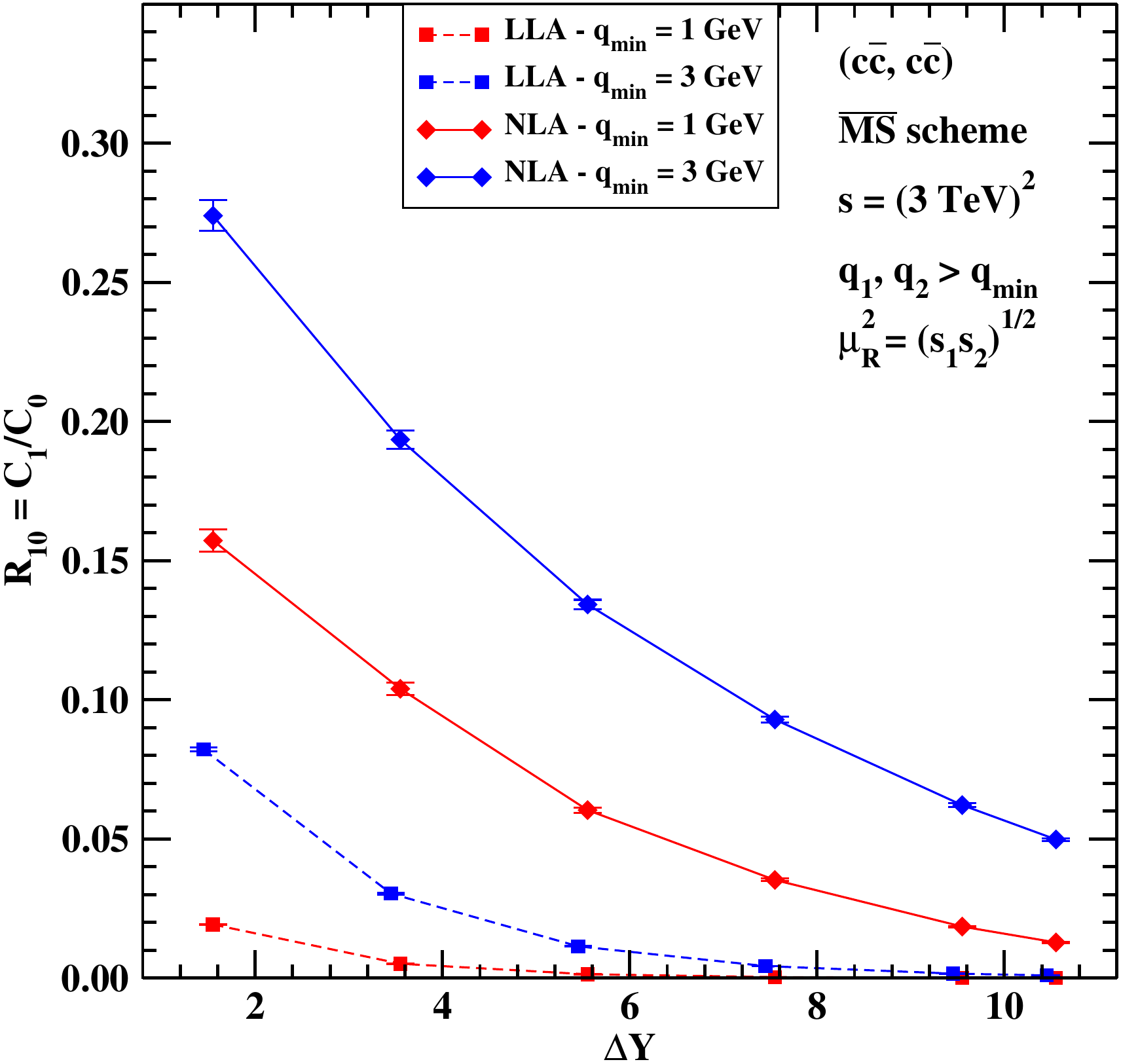}

   \caption{$\Delta Y$-dependence of $C_0$ and $R_{10}$ for $q_{\rm min} = 1$,
     3 GeV, $\sqrt{s} = 3$ TeV, and for $\mu_R^2 = \sqrt{s_1s_2}$, with
     $s_{1,2} = m_{1,2}^2 + q_{1,2}^2$. Data points have been slightly shifted
     along the horizontal axis for the sake of readability.}
\label{fig:3TeV}
\end{figure}

Results for $C_0$, $R_{10}$, and $R_{20}$ for $q_{\rm min} = 1$, 3 GeV and
$\sqrt{s} = 200$ GeV are shown in Fig.~\ref{fig:200GeV}. For comparison,
predictions for $C_0$ and $R_{10}$ for $q_{\rm min} = 1$, 3 GeV and
$\sqrt{s} = 3$ TeV are shown in Fig.~\ref{fig:3TeV}.

\begin{table}[tb]
\centering
\caption{$\Delta Y$-dependence of the $\varphi$-averaged cross section
  $C_0$ [pb] for $q_{\rm min} = 0$ GeV and $\sqrt{s} = 200$ GeV. LLA and NLA
  predictions are compared with the ``box'' $q\bar{q}$ cross section.
  $C$ stands for $\mu_R^2/(s_1 s_2)$.}
\label{tab:C0-200GeV}
\begin{tabular}{r|lllllll}
\toprule
$\Delta Y$ & 
Box $q\bar{q}$ &
$\barr c \rm LLA \\ C = 1/2 \earr$ & 
$\barr c \rm LLA \\ C = 1 \earr$ &
$\barr c \rm LLA \\ C = 2 \earr$ &
$\barr c \rm NLA \\ C = 1/2 \earr$ &
$\barr c \rm NLA \\ C = 1 \earr$ & 
$\barr c \rm NLA \\ C = 2 \earr$ \\
\midrule
1.5 & 98.26 & 51.87(17) & 8.155(39) & 3.618(18) & 2.120(13) & 1.4046(91) & 1.2861(93) \\
2.5 & 42.73 & 90.46(26) & 11.080(44) & 4.322(21) & 2.197(11) & 1.1976(71) & 1.0623(70) \\
3.5 & 14.077 & 150.42(43) & 14.166(54) & 4.876(20) & 2.315(12) & 0.9986(54) & 0.8296(45) \\       
4.5 & 3.9497 & 231.45(62) & 16.705(65) & 5.053(24) & 2.301(11) & 0.7763(39) & 0.6116(32) \\
5.5 & 0.9862 & 319.93(89) & 17.529(69) & 4.648(21) & 2.121(10) & 0.5411(27) & 0.3922(19) \\
\bottomrule
\end{tabular}
\end{table}

\begin{table}[tb]
\centering
\caption{$\Delta Y$-dependence of the $\varphi$-averaged cross section
  $C_0$ [pb] for $q_{\rm min} = 0$ GeV and $\sqrt{s} = 3$ TeV. LLA and NLA
  predictions are compared with box $q\bar{q}$ cross section.
  $C$ stands for $\mu_R^2/(s_1 s_2)$.}
\label{tab:C0-3TeV}
\setlength{\tabcolsep}{5.3pt}
\begin{tabular}{r|lllllll}
\toprule
$\Delta Y$ & 
Box $q\bar{q}$ &
$\barr c \rm LLA \\ C = 1/2 \earr$ & 
$\barr c \rm LLA \\ C = 1 \earr$ &
$\barr c \rm LLA \\ C = 2 \earr$ &
$\barr c \rm NLA \\ C = 1/2 \earr$ &
$\barr c \rm NLA \\ C = 1 \earr$ & 
$\barr c \rm NLA \\ C = 2 \earr$ \\
\midrule
1.5 & 280.98 & 1361.6(6.1) & 66.40(30) & 24.44(11) & 12.45(11) & 7.292(72) & 6.521(73) \\
3.5 & 48.93 & 6856(18) & 196.12(95) & 54.93(26) & 23.07(14) & 8.153(62) & 6.798(59) \\
5.5 & 4.9819 & 31860(71) & 551.2(2.4) & 116.33(53) &  47.53(23) & 9.479(67) & 6.903(45) \\       
7.5 & 0.4318 & 130215(271) & 1365.1(5.5) & 217.9(1.0) & 94.54(44) & 10.243(56) & 6.435(33) \\
9.5 & 0.0323 & 428626(977) & 2691.2(9.9) & 327.3(1.5) & 158.38(76) & 9.092(45) & 4.858(24) \\
10.5& 0.0081 & 683469(1833) & 3278(13) & 345.2(1.5) & 180.42(90) & 7.497(37) & 3.651(18) \\
\bottomrule
\end{tabular}
\end{table}

%
%

\subsection{Numerical tools and uncertainty estimation}

All numerical calculations were done in \textsc{Fortran}. Numerical integrations
were performed using routines implemented in the {\tt Cuba}
library~\cite{Cuba:2005,ConcCuba:2015}, making extensive use of the Monte
Carlo {\tt Vegas}~\cite{VegasLepage:1978} integrator. The numerical stability
of our results was crosschecked by separate calculations performed using
both \textsc{Mathematica} and the {\tt Dadmul} CERNLIB routine~\cite{cernlib}.

The most important source of uncertainty comes from the numerical
six-dimensional integration over the variables $|\vec q_1|$, $|\vec q_2|$,
$y_1$, $\nu$, $x_1$, and $x_2$ and was directly estimated by the {\tt Vegas}
integration routine~\cite{VegasLepage:1978}. 
We checked that other sources of uncertainties, related with the upper cutoff
in the integrations over $|\vec q_1|$, $|\vec q_2|$, and $\nu$, are negligible
with respect to the first one. Thus, the error bars of all predictions are just
those given by {\tt Vegas}. The other, internal source of uncertainty of our
calculation is related with the scale of the running QCD coupling. Below we
quantify this uncertainty studying the renormalization scale
dependence.  We vary $\mu_R^2$ around its ``natural'' value
$\sqrt{s_1 s_2}$ in the range 1/2 to two. The parameter $C$ entering Tables~\ref{tab:C0-200GeV}
and~\ref{tab:C0-3TeV} gives the ratio $\mu_R^2/\sqrt{s_1 s_2}$.

\subsection{Discussion}

The inspection of results in Table~\ref{tab:C0-200GeV} suggests that the cross
section $C_0$ is smaller than the reference ``box'' cross section.
This is similar to what occurred in the calculations of the total
$\gamma^*\gamma^*$ cross section,
where it was found that the ``box'' mechanism gives still a very important
contribution at LEP2 energies. The situation changes if we pass to the larger
energies and therefore larger rapidity differences that are possibly available
at future $e^+e^-$ colliders. Here the BFKL mechanism with the gluonic exchange
in the $t$-channel starts to win over the ``box'' one with the fermionic
$t$-channel exchange, see our results in Table~\ref{tab:C0-3TeV}. 
We stress, however, that for our two heavy-quark (or two heavy-antiquark)
tagged process, contrary to the $\gamma^*\gamma^*$ case, the ``box'' mechanism
is not a background.
The cross section exhibits the expected trends: it increases when moving from
the LEP2 energies to the CLIC ones and decreases when moving from the LLA to
the NLA, a typical feature in the BFKL approach. We note also that our partial
inclusion of NLA effects leads to results that are less sensitive to the
variation of the renormalization scale than the results of LLA BFKL
resummation.

Azimuthal correlations are in all cases much smaller than one and decrease
when $\Delta Y$ increases, as it must be due to the larger emission of
undetected partons. The reason for this smallness, with respect to the
case of Mueller-Navelet jets or di-hadron systems (see, {\it e.g.},
Refs.~\cite{Caporale:2012ih,Celiberto:2017ptm}) is that in this case there is
not any kinematic constraint, even at the lowest order in perturbation theory,
between the transverse momenta of the two tagged quarks, since they are
produced in two different vertices (each of them together with an antiquark).
When the minimum value of the tagged quark transverse momentum $q_{\rm min}$
is increased, azimuthal correlations increase due to the more limited
available phase space in the transverse space and the consequently more
constrained transverse kinematics. We can see that the inclusion of NLA
effects increases the correlations, which can only be explained with the
larger suppression of $C_0$ with respect to $C_{1,2}$ when these effects are
included.  

\section{Summary and outlook}

We have considered the inclusive photoproduction of two heavy quarks
separated in rapidity, taking into account the resummation to all orders
of the leading energy logarithms and the resummation of the next-to-leading
ones entering the BFKL Green's function. We have calculated the cross section
for this process averaged over the relative azimuthal angle of the two tagged
quarks and presented results for the azimuthal angle correlations, considering
for definiteness the case
in which photons are emitted by electron and positrons colliding at the energies
of the LEP2 and the CLIC colliders.

The trends of our results with the energy
of the collision beam and the behavior of the considered observables with
the rapidity interval $\Delta Y$ between the tagged quarks is just as expected:
azimuthal correlations decrease with increasing $\Delta Y$. Moreover, just
as in the case of Mueller-Navelet jets and dihadron production, the inclusion
of next-to-leading order correction reduces the decorrelation. In absolute
value, azimuthal correlations are much smaller with respect to Mueller-Navelet
jets and dihadrons, a result which is not surprising since here we have a two
heavy-quark pair production mechanism and the two tagged quarks are produced
in the leading order in different interaction vertices, having independent
transverse momenta.

This process extends the list of semihard processes by which strong interactions
in the high-energy limit, and in particular the BFKL resummation procedure,
can be probed in the future $e^+e^-$ linear colliders.

There are several obvious developments of this work. One is the calculation of
the next-to-leading order impact factor for the photoproduction of a heavy
quark-antiquark pair, which would allow for the full NLA treatment of the
process under consideration. The other is to include into the theoretical
analysis heavy-quark fragmentation describing the experimental tagging
procedure of heavy quarks. 

As we already noted, the possibility for the experimental study of our process
in UPC collisions of heavy ions at the LHC kinematics is not feasible,
unfortunately. Nevertheless, the study of heavy-quark observables that reveal
BFKL resummation effects looks promising in the LHC proton-proton collisions.
Recently, in~\cite{Boussarie:2017oae} the process of inclusive forward
$J/\Psi$-meson and very backward jet production was suggested. 
The other interesting possibility is to extend the methods used in our work   
to the case of two detected heavy-quark inclusive hadroproduction, {\it i.e.}
a process similar to the one considered here, but
initiated by quarks and gluons emitted by protons (in collinear factorization)
rather than by photons.

\section*{Acknowledgements}

F.G.C. acknowledges support from the Italian Foundation ``Angelo della Riccia'' and thanks the Instituto de F\'isica Te\'orica (IFT UAM-CSIC) in Madrid for warm hospitality.
\\
D.I. thanks the Dipartimento di Fisica dell'U\-ni\-ver\-si\-t\`a della Calabria
and the Istituto Nazio\-na\-le di Fisica Nucleare (INFN), Gruppo collegato di
Cosenza, for the warm hospitality and the financial support. The work
of D.I. was also supported in part by the Russian Foundation for Basic Research
via grant RFBR-15-02-05868.
\\
The work of B.M. was supported by the European Commission, European Social Fund
and Calabria Region, that disclaim any liability for the use that can be done
of the information provided in this paper.
B.M. thanks the Sobolev Institute of Mathematics of Novosibirsk for warm hospitality during the preparation of this work.

\end{document}